# Desain dan Implementasi *Face Recognition* dan *Live Streaming* pada Sistem *Digital Assistant* untuk Staf Medik Fungsional Menggunakan Google Glass

Muhammad Nur Pratama[*1] dan Ary Setijadi Prihatmanto[*2]

*Program Studi Teknik Elektro*

*Sekolah Teknik Elektro dan Informatika*

*Institut Teknologi Bandung, Jl. Ganesha No. 10, Bandung 40132, Indonesia*

[1]`mnurpratama@outlook.com`

[2]`asetijadi@lskk.ee.itb.ac.id`

*Abstrak*— **Dalam era globalisasi saat ini, rumah sakit dituntut untuk meningkatkan kinerja dan daya saing sebagai badan usaha dengan tidak mengurangi misi sosial yang dibawanya. Hal ini berarti bahwa rumah sakit harus menerapkan kebijakan-kebijakan strategis agar mampu secara cepat dan tepat dalam pengambilan keputusan sehingga dapat menjadi organisasi yang responsif, inovatif, efektif, dan efisien. DIANE merupakan sebuah sistem *digital assistant* yang bertujuan untuk mempercepat akses staf medik fungsional sebagai seorang profesional dengan waktu terbatas terhadap informasi seperti layanan kesehatan, rekam medis, dan personalia pada SIRS tingkat rumah sakit. Dengan menggunakan *face recognition*, DIANE dapat mengidentifikasi pasien sehingga dapat memberikan tampilan data berupa biodata dan rekam medis pasien. Sedangkan menggunakan *live streaming* pada DIANE, seorang staf medik fungsional dapat berkomunikasi dengan staf medik fungsional di tempat lain. Dengan peningkatan kecepatan akses terhadap beragam informasi tersebut diharapkan seorang staf medik fungsional dapat mengambil keputusan dengan lebih cepat. Akibatnya, semakin banyak pasien yang dapat ditangani oleh staf medik fungsional dengan baik.**

*Kata Kunci*— ***digital assistant*, staf medik fungsional, google glass.**

## I. PENDAHULUAN

Dalam era globalisasi saat ini, rumah sakit dituntut untuk meningkatkan kinerja dan daya saing sebagai badan usaha dengan tidak mengurangi misi sosial yang dibawanya. Hal ini berarti bahwa rumah sakit harus menerapkan kebijakan-kebijakan strategis dalam hal peningkatan kinerja dan pelayanan, efisiensi dari dalam (organisasi, manajemen, dan sumber daya manusia) serta harus mampu secara cepat dan tepat dalam pengambilan keputusan agar dapat menjadi organisasi yang responsif, inovatif, efektif, dan efisien. Salah satu kebijakan strategis yang dapat diambil adalah pemanfaatan teknologi informasi pada SIRS (Sistem Informasi Rumah Sakit).

Selain SIRS, kualitas pelayanan rumah sakit sangat dipengaruhi oleh kinerja sumber daya rumah sakit, salah satunya adalah staf medik fungsional. Staf medik fungsional bersifat dinamis dan memiliki berbagai tugas seperti melakukan praktik, kunjungan ke pasien rawat inap, dan tindakan operasi. Oleh karena itu, staf medik fungsional lebih membutuhkan keberadaan suatu sistem yang dapat mempercepat akses terhadap informasi seperti layanan kesehatan, rekam medis, dan personalia pada SIRS tingkat rumah sakit. Dengan menggunakan sistem tersebut, staf medik fungsional sebagai seorang profesional dengan waktu terbatas dapat meningkatkan produktivitas kerja dan kinerjanya.

Berangkat dari masalah-masalah tersebut, dibuat sebuah sistem *digital assistant* yang bertujuan untuk mempercepat akses staf medik fungsional sebagai seorang profesional dengan waktu terbatas terhadap informasi seperti layanan kesehatan, rekam medis, dan personalia pada SIRS tingkat rumah sakit. Peningkatan kecepatan akses staf medik fungsional terhadap informasi pada SIRS tingkat rumah sakit dapat dilakukan melalui penggunaan Google Glass yang merupakan *wearable device* berjenis OHMD (*Optical Head-Mounted Display*). Dengan menggunakan Google Glass, staf medik fungsional dapat mengakses informasi pada SIRS tingkat rumah sakit dan menangani pasien dalam waktu bersamaan.

## II. STUDI PUSTAKA

### A. *SIRS (Sistem Informasi Rumah Sakit)*

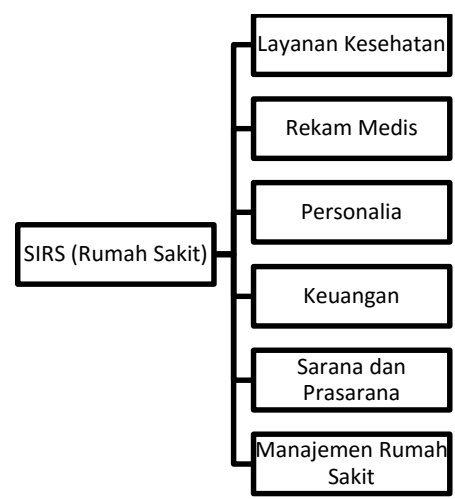

**Gambar 1** SIRS Tingkat Rumah Sakit

SIRS merupakan sebuah sistem informasi rumah sakit terintegrasi yang terdiri dari tiga tingkat yaitu tingkat pemerintah pusat, tingkat pemerintah daerah, dan tingkat rumah sakit. Pada tingkat rumah sakit, SIRS disusun dengan menggunakan pendekatan fungsi dan dilakukan secara menyeluruh. Akibatnya, SIRS tingkat rumah sakit terdiri atas layanan kesehatan, rekam medis, personalia, keuangan, sarana dan prasarana, dan manajemen rumah sakit.

### B. GOOGLE GLASS

Google Glass merupakan sebuah perangkat OHMD yang berbentuk seperti kacamata. Perangkat tersebut dikembangkan dengan tujuan untuk mempermudah akses manusia terhadap sebuah komputer. Google Glass mampu memberikan pengalaman pengguna seperti sebuah ponsel pintar. Hal tersebut disebabkan Google Glass menggunakan sistem operasi berbasis Android. Penggunanya dapat mengoperasikan Google Glass melalui sentuhan atau perintah suara. Google Glass dipublikasikan pertama kali pada tanggal 15 April 2013 di Amerika dengan harga $1,500. Namun, Google Glass baru diperjualbelikan secara komersil terhitung tanggal 15 Mei 2014.

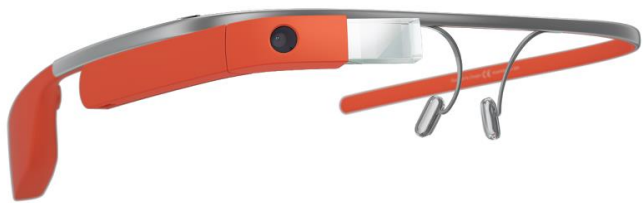

**Gambar 2** Perangkat Keras Google Glass

### C. SISTEM OPERASI ANDROID

Android adalah sistem operasi berbasis Linux yang dirancang untuk beberapa perangkat seluler seperti ponsel pintar, tablet, dan *wearable device* seperti Google Glass. Sistem operasi ini dirilis secara resmi pada tahun 2007. Android adalah sistem operasi *open source* sehingga memungkinkan untuk dilakukannya modifikasi secara bebas. Fitur-fitur pada Android umumnya ditulis dalam bahasa pemograman Java.

Untuk dapat membuat perangkat lunak berbasis Android, umumnya digunakan alat bantu pengembangan berupa perangkat lunak IDE (*Integrated Development Environment*) Android Studio. Di dalam Android Studio terdapat seperangkat alat bantu seperti *debugger*, *library*, *emulator*, dokumentasi, contoh kode, dan tutorial. Dalam pengembangan perangkat lunak berbasis Android pada Google Glass turut dibutuhkan *library* berupa GDK (*Glass Development Kit*).

### D. DATABASE

Database atau basis data adalah kumpulan informasi yang disimpan di dalam komputer secara sistematik sehingga dapat diperiksa menggunakan suatu perangkat lunak komputer untuk memperoleh informasi dari basis data tersebut.

### E. WEB SERVICE

Agar sebuah perangkat lunak pada sistem operasi Android dapat memanipulasi data yang tersimpan pada database, diperlukan penghubung antara perangkat lunak dengan database berupa web service. Web service menyediakan autentikasi database sehingga seluruh perangkat lunak yang terhubung dengan web service dapat memanipulasi data yang tersimpan pada database.

Hingga saat ini terdapat dua jenis *communication protocol* yang dapat menghubungkan antara perangkat lunak dengan web service, yaitu SOAP (*Simple Object Access Protocol*) dan REST (*Representational State Transfer*). Keduanya memiliki kelebihan dan kekurangannya masing-masing bergantung pada kebutuhan pengguna.

### F. FACE RECOGNITION

*Face recognition* merupakan sebuah metode pengolahan citra digital untuk mengidentifikasi seseorang menggunakan masukan berupa citra wajah. Saat ini, terdapat beberapa teknik pengolahan citra digital yang cukup terkenal dan dapat digunakan untuk mengidentifikasi identitas pemilik citra wajah seperti *eigen classifier*, *fisher classifier*, dan *local binary pattern histogram classifier*.

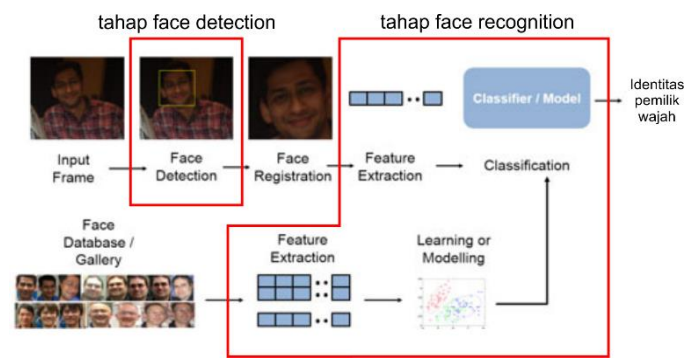

**Gambar 3** Mekanisme Kerja *Face Recognition*

### G. LIVE STREAMING

*Live streaming* merupakan sebuah metode untuk dapat mentransmisikan citra dari kamera sebuah perangkat keras secara *realtime* ke server. Citra *realtime* yang diterima oleh server turut dapat diakses dan diunduh oleh pengguna lain.

## III. PERANCANGAN PRODUK

### A. PERANCANGAN PERANGKAT KERAS

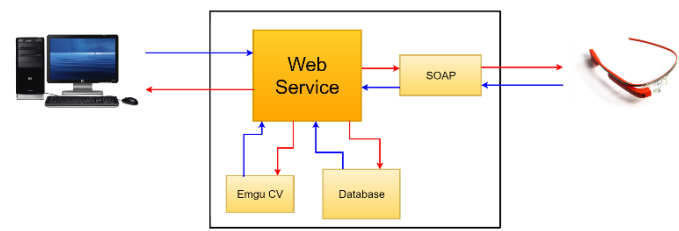

**Gambar 4** Perancangan Perangkat Keras

Sistem ini dibentuk berdasarkan jaringan berarsitektur *client-server*. Sistem ini terdiri dari beberapa perangkat keras seperti Google Glass sebagai *client* dan PC sebagai server. Google Glass dapat terhubung dengan server melalui perantara perangkat jaringan. Pada sistem ini, server berperan sebagai media penyimpan keseluruhan informasi seperti layanan kesehatan, rekam medis, dan personalia dalam bentuk database SIRS tingkat rumah sakit. Agar Google Glass dan server dapat terhubung, maka dibutuhkan web service sebagai pemberi autentikasi sehingga Google Glass dapat terhubung dengan server.

### B. PERANGKAT LUNAK PADA GOOGLE GLASS

Berdasarkan masalah-masalah yang muncul kemudian dirancang sebuah perangkat lunak untuk dijalankan pada Google Glass. Spesifikasi perangkat lunak pada Google Glass adalah sebagai berikut:

1. *Face recognition* untuk mengidentifikasi pasien
2. Setelah pasien dikenali, proses akuisisi data pasien berupa biodata dan rekam medis dari database SIRS tingkat rumah sakit pada server dapat dijalankan. Biodata dan rekam medis pasien ditampilkan pada layar Google Glass
3. Komunikasi antarpengguna dengan menggunakan *live streaming*

Berdasarkan spesifikasi perangkat lunak tersebut selanjutnya dilakukan perancangan berupa modul-modul penyusun perangkat lunak di Google Glass. Modul-modul tersebut terlihat seperti pada gambar berikut.

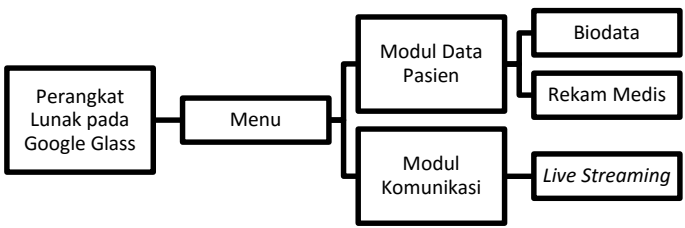

Gambar 5 Komponen Penyusun Perangkat Lunak pada Google Glass

Melalui Gambar 5, selanjutnya dapat disusun skenario penggunaan perangkat lunak pada Google Glass. Skenario tersebut ditunjukkan melalui diagram seperti pada gambar berikut.

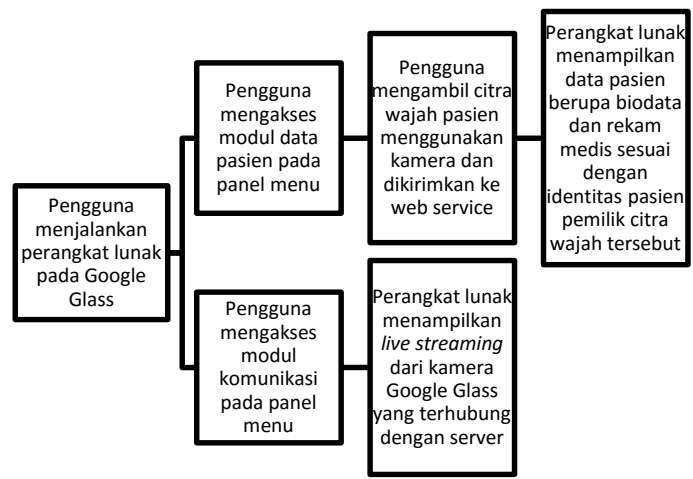

Gambar 6 Skenario Penggunaan Perangkat Lunak pada Google Glass

Seperti terlihat pada Gambar 6 bahwa untuk mendapatkan data pasien berupa biodata dan rekam medis pada modul data pasien, pengguna diminta untuk mengirimkan citra wajah pasien ke web service. Citra wajah tersebut selanjutnya akan menjalani proses *face recognition* di web service agar identitas pasien pemilik citra wajah dapat dikenali. Alur data modul data pasien di perangkat lunak pada Google Glass dapat dilihat melalui *data flow diagram* seperti pada gambar berikut.

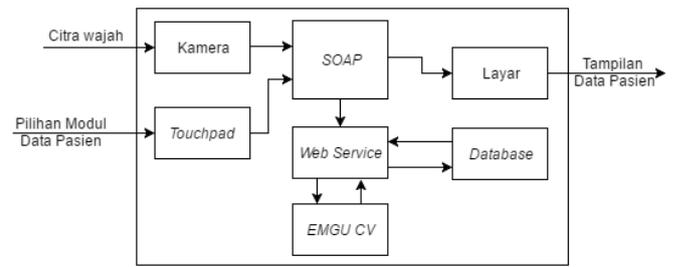

Gambar 7 *Data Flow Diagram* Modul Data Pasien pada Perangkat Lunak di Google Glass

Selain modul data pasien, pada perangkat lunak di Google Glass turut terdapat modul komunikasi berupa *live streaming* yang memungkinkan pengguna di tempat lain dapat menyaksikan pemindaian citra dari kamera Google Glass secara *realtime*. Alur data modul komunikasi di perangkat lunak pada Google Glass dapat dilihat melalui *data flow diagram* seperti pada gambar berikut.

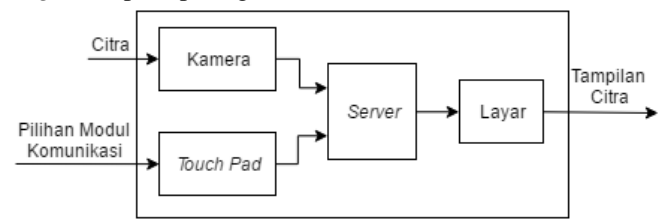

Gambar 8 *Data Flow Diagram* Modul Komunikasi pada Perangkat Lunak di Google Glass

### C. WEB SERVICE PADA SERVER

Agar fitur *face recognition* pada modul data pasien di perangkat lunak pada Google Glass dapat dijalankan, maka dibutuhkan perancangan proses identifikasi citra wajah yang dilakukan di web service pada server.

Web service yang digunakan pada sistem ini adalah berbasis *communication protocol* SOAP. Hal tersebut disebabkan oleh kemudahan perancangan sistem jika dibandingkan dengan *communication protocol* lain, seperti REST. SOAP juga memberikan kemudahan dalam pengembangan *multi platform*, artinya web service berbasis SOAP dapat digunakan tidak hanya oleh perangkat lunak berbasis Android.

Pada modul *face recognition* di web service terdapat dua tahapan, yaitu tahap *face detection* dan tahap *face recognition*. Kedua tahapan tersebut dibangun dengan menggunakan konstruksi OpenCV.

Pada tahap *face detection* dilakukan pendeteksian keberadaan wajah dari sebuah citra hasil tangkapan kamera. Pendeteksian dilakukan dengan menggunakan teknik yang bernama *template matching*. Pada teknik *template matching* dilakukan pencocokan antara citra hasil tangkapan kamera dengan template citra wajah. Template citra wajah yang dimaksud adalah berupa komponen penyusun wajah seperti mata, hidung, dan mulut. Proses pendeteksian wajah dengan menggunakan teknik *template matching* tersebut biasanya turut dikenal sebagai metode Viola-Jones (atau dikenal juga sebagai metode Haar cascades).

Citra wajah hasil proses dari tahap *face detection* selanjutnya diproses pada tahap *face recognition*. Tahap *face recognition* pada modul *face recognition* di web service pada server menggunakan metode *eigenfaces classifier*.

Pada tahap *face recognition* dilakukan pembandingan antara citra wajah yang didapat dengan berbagai citra wajah yang telah tersimpan di database. Melalui proses pembandingan tersebut dapat diketahui tingkat kemiripan antara citra wajah yang didapat dengan berbagai citra wajah yang tersimpan di database. Jika tingkat kemiripan melebihi suatu nilai *threshold* yang diatur, maka modul *face recognition* akan mengembalikan data berupa informasi dari pemilik citra wajah tersebut ke modul data pasien pada perangkat lunak di Google Glass. Sedangkan jika tingkat kemiripan tidak melebihi suatu nilai *threshold* yang diatur, maka modul *face recognition* akan mengembalikan informasi berupa identitas pemilik citra wajah tersebut tidak berhasil dikenali.

Keseluruhan proses pada modul *face recognition* di web service pada server dapat dilihat melalui diagram alir seperti pada gambar berikut.

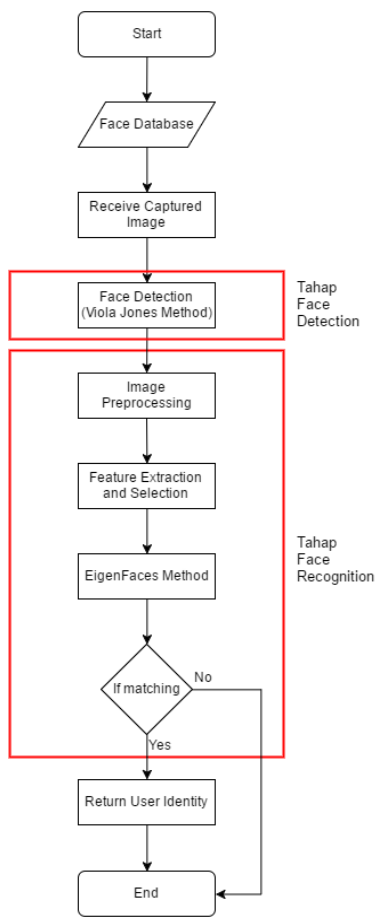

**Gambar 9** Diagram Alir Modul *Face Recognition* pada Web Service di Server

IV. IMPLEMENTASI PRODUK

A. *Perangkat Lunak pada Google Glass*

Setelah modul data pasien diaktifkan, maka munculah tampilan berupa hasil pindai citra dari kamera Google Glass. Pada tampilan tersebut akan terlihat garis kotak berwarna kuning di bagian tengah layar dan tulisan "*click on touchpad to capture*" di bagian bawah layar. Agar proses selanjutnya dapat dilanjutkan, pengguna diminta untuk memberi masukan berupa sentuhan melalui *touchpad* ketika wajah pasien tepat berada di dalam kotak berwarna kuning.

Ketika masukan berupa sentuhan berhasil diterima oleh Google Glass, maka perangkat lunak akan mengirimkan hasil pindai citra berisikan wajah pasien ke modul *face recognition* pada web service di server. Pengiriman citra wajah pasien ke web service dilakukan melalui *communication protocol* SOAP menggunakan jaringan internet. Hasil proses pada web service yang berupa data pasien atau informasi bahwa citra wajah tidak berhasil diidentifikasi akan diterima oleh Google Glass sesaat setelah proses identifikasi citra wajah pada modul *face recognition* di web service selesai dijalankan.

Jika proses identifikasi citra wajah pasien berhasil dilakukan, maka web service akan mengirimkan kembali data pasien berupa biodata dan rekam medis ke perangkat lunak pada Google Glass. Biodata pasien yang dimaksud adalah berupa citra wajah *profile picture* pasien untuk proses verifikasi, nama pasien, jenis kelamin pasien, dan tanggal lahir pasien. Sedangkan data pasien berupa rekam medis yang dimaksud adalah berupa alergi yang pernah diderita oleh pasien, imunisasi yang pernah diberikan ke pasien, dan obat-obatan yang pernah diberikan ke pasien. Penjelasan mengenai tampilan modul data pasien adalah terlihat seperti pada tabel berikut.

**Tabel 1** Modul Data Pasien pada Perangkat Lunak di Google Glass

| Tampilan pindai citra dari kamera Google Glass | Pemrosesan citra wajah pasien di modul *face recognition* pada web service di server (setelah *touchpad* Google Glass disentuh) | Tampilan biodata pasien (halaman pertama) |
|---|---|---|
| | | Yusrina Nur Dini, Female, Born on 2/2/1994, Patient ID 8 |
| | | Tampilan alergi yang pernah diderita pasien (halaman kedua) |
| | | Latex 3/3/2016, Kacang 5/18/2016, Debu 6/8/2016, Allergies Yusrina Nur Dini |
| | | Tampilan imunisasi yang pernah diberikan ke pasien (halaman ketiga) |
| | | Meningitis 4/3/2016, Vaksin Kanker 5/18/2016, Hepatitis A 6/8/2016, Immunization Yusrina Nur Dini |

Selain modul data pasien, pada perangkat lunak di Google Glass turut terdapat modul komunikasi berupa *live streaming*. Setelah modul komunikasi diaktifkan, maka munculah tampilan berupa hasil pindai citra dari kamera pada layar Google Glass. Hasil pindai citra yang didapatkan selanjutnya dikirim ke server agar pengguna sistem yang lain turut dapat menyaksikan kejadian yang terekam oleh kamera Google Glass.

Layanan *live streaming* pada modul komunikasi di perangkat lunak pada Google Glass dibangun berdasarkan konstruksi kickflip. Kickflip menyediakan SDK (*Software Development Kit*) yang dapat digunakan pada pengembangan perangkat lunak berbasis Android. Selain SDK, kickflip juga menyediakan layanan berupa server yang dapat digunakan sebagai media penyimpan citra berupa video hasil *live streaming*.

Penjelasan mengenai mekanisme kerja dari modul komunikasi pada perangkat lunak di Google Glass adalah terlihat seperti pada tabel berikut.

**Tabel 2** Modul Komunikasi pada Perangkat Lunak di Google Glass

| Citra hasil pindai kamera Google Glass ditampilkan pada layar Google Glass. Selain itu, citra hasil pindai kamera Google Glass turut dikirimkan ke server kickflip melalui jaringan internet agar dapat diunduh oleh pengguna lain melalui website kickflip. |
|---|
| 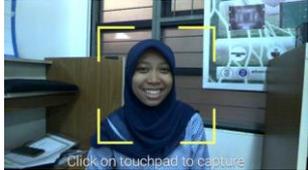 |
| Melalui website kickflip, seorang pengguna lain dapat mengunduh citra hasil pindai kamera Google Glass secara *realtime*. Proses tersebut hanya dapat dilakukan saat modul komunikasi pada perangkat lunak di Google Glass sedang dijalankan. |
| 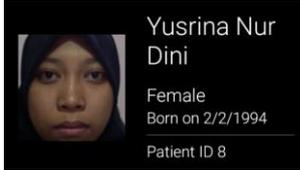 |

### B. WEB SERVICE PADA SERVER

Implementasi web service pada server dilakukan berdasarkan perancangan produk yang telah dilakukan. Pada pemrograman web service pada server, sebuah fitur atau web method dapat terdiri dari berbagai komponen penyusun yang berupa file dengan ekstensi .ASMX dan kode pendukung dengan bahasa pemrograman C#. Hal tersebut dikarenakan pada sistem ini digunakan pemrograman web service berbasis *communication protocol* SOAP. Perangkat lunak IDE yang digunakan untuk mengembangkan web service adalah Microsoft Visual Studio.

Modul *face recognition* pada web service di server diimplementasikan sesuai dengan diagram alir pada Gambar 9. Hasil yang didapatkan adalah berupa proses *face recognition* yang berlangsung seperti terlihat pada tabel berikut.

| Hasil pindai citra dari kamera Google Glass yang dikirimkan ke modul *face recognition* pada web service | Proses identifikasi citra wajah yang berhasil dilakukan. Hal tersebut dibuktikan dari kemiripan antara *profile picture* seperti pada gambar di bawah dengan masukan hasil pindai citra dari kamera Google Glass. |
|---|---|
| 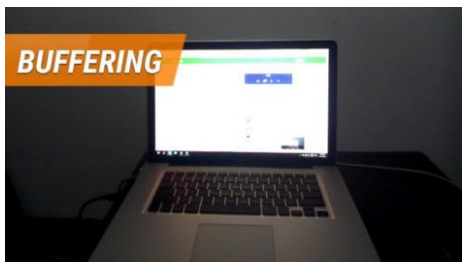 | 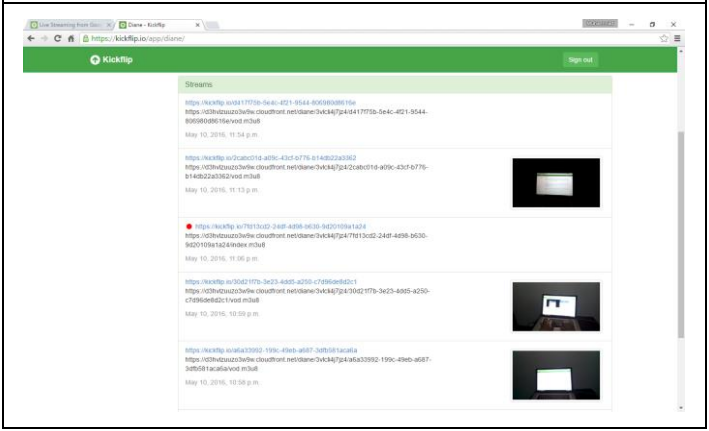 |

### V. KESIMPULAN

Kesimpulan yang dapat diambil setelah dilakukan implementasi produk adalah sebagai berikut:

1. Perangkat lunak pada Google Glass dapat berkomunikasi dengan server melalui web service. Komunikasi yang terjalin antara perangkat lunak pada Google Glass dengan web service adalah berupa pengunggahan dan pengunduhan data.
2. Pengoperasian antarmuka perangkat lunak pada Google Glass disesuaikan dengan arahan desain dari Google serta kebutuhan pengguna. Berdasarkan pengujian yang telah dilakukan didapatkan kesesuaian antara hasil implementasi antarmuka perangkat lunak dengan rancangan pada dokumen B300.
3. Fitur *face recognition* berhasil diimplementasikan pada perangkat lunak di Google Glass untuk mengidentifikasi pasien sehingga selanjutnya perangkat lunak tersebut dapat menampilkan data pasien berupa biodata dan rekam medis pasien. Proses implementasi *face recognition* dilakukan dengan dukungan dari OpenCV berupa penyediaan SDK pengolahan citra digital.
4. Fitur *live streaming* berhasil diimplementasikan pada perangkat lunak di Google Glass sehingga pengguna lain dapat mengakses atau mengunduh citra dari kamera Google Glass secara *realtime*. Proses implementasi *live streaming* dilakukan dengan dukungan dari kickflip berupa penyediaan SDK dan server.

### DAFTAR PUSTAKA

[1] D. Tychalas and A. Kakarountas, *Planning and Development of an Electronic Health Record Client Based on the Android Platform*, Informatics (PCI), 2010 14th Panhellenic Conference on, Tripoli, 2010, pp. 3-6.
[2] T. H. Bloebaum and F. T. Johnsen, *Exploring SOAP and REST communication on the Android platform*, Military Communications Conference, MILCOM 2015 - 2015 IEEE, Tampa, FL, 2015, pp. 599-604.
[3] Tayal, Yogesh et al., *Face Recognition using Eigenface*, 3rd Ed., Department of Electrical and Instrumentation Engineering, Institute of Technology and Management Gwalior M. P., 2013.
[4] Turk, Matthew et al., *Eigenfaces for Recognition*, 1st Ed., Massachusetts Institute of Technology, 1991.